\documentclass[reprint,amsmath,amssymb,aps]{revtex4-2}

\usepackage{hyperref}
\hypersetup{
    colorlinks=true,
    linkcolor=blue,
    citecolor=blue,
    urlcolor=blue
}

\usepackage{graphicx}

\usepackage{mathdots}
\usepackage{xcolor}
\usepackage{xspace}

\usepackage[toc,page]{appendix}

\begin{document}

\title{Observation of temporal Wood's anomaly in folded time gratings: surface-wave-enhanced transmission and the emergence of gain}

\author{Amit Shaham}
\email{samitsh@campus.technion.ac.il}

\author{Ben-Zion Joselson}

\author{Ilya Varenisov}

\author{Denis Dikarov}

\author{Ariel Epstein}
\email{epsteina@ee.technion.ac.il}

\affiliation{Andrew and Erna Viterbi Faculty of Electrical and Computer Engineering, Technion -- Israel Institute of Technology, Haifa 3200003, Israel}

\begin{abstract}
{Wood's anomaly is a fundamental wave phenomenon that stems from the interplay between farfields and surface-wave (SW) resonances through structured interfaces. Recent theories have suggested temporal analogs of such SW coupling processes by employing frequency transitions via time-periodic interfaces, rather than classical wavevector (momentum) transitions via space-periodic gratings. In this paper, we observe this phenomenon experimentally by devising a folded time grating of a single, waveguide-enclosed, time-modulated element; this substantially reduces complexity and power consumption and facilitates transmissive operation. We support our findings by deriving a comprehensive Floquet-Bloch analysis that exhibits excellent agreement with measurements. Importantly, we utilize our framework and experiment to reveal a unique regime of temporal Wood's anomaly, in which coupling to negative SW frequencies manifests tunable parametric amplification unattainable via traditional spatial modulation. Beyond the fundamental contribution, our results provide an economical path for universally synthesizing intricate temporal apertures to enhance dynamic filtering and leaky-wave antennas.
}
\end{abstract}

{\hypersetup{hidelinks}
\maketitle
}

\section*{Introduction}
\label{Sec:Intro}
The traditional way to control light-matter interactions to the extreme would be to geometrically structure materials in space, for instance, to form metamaterials \cite{Engheta2006,Capolino2009,Zheludev2012}. In such artificial media, the macroscopic wave properties are directly formed by microscopic scattering off miniature inclusions, which is manifested through wavevector (momentum) exchange, while the frequency (energy) remains conserved.

The fundamentals of such momentum-transition processes can be drawn to a great extent from the elementary class of periodic structures \cite{Brillouin1953,Oliner1959,Tamir1964,Hessel1973,Elachi1976}. As an important archetype, two-dimensional diffraction gratings scatter incident light into a discrete set of Floquet-Bloch (FB) harmonics, owing to the transverse momenta contributed by the lattice \cite{Loewen1997,Palmer2002}.
However, beneath this simplistic picture of diffraction lies a more intricate electromagnetic mechanism of interelement coupling. Particularly, collective resonances form surface- (SW) or leaky- (LW) wave eigenmodes that, in principle, sustain themselves without external excitation \cite{Fano1941,Oliner1959,Hessel1965}.

Tuned via the geometric features of the grating, the dispersion of such self-resonances plays a central role in determining its scattering behavior. The grating couples the excitation to other spatial harmonics and back, and thus imprints part of their response onto the fundamental harmonic. Notably, if the momentum associated with one of these harmonics resides near an eigenmode pole, abrupt variations typically appear in the intensity of the scattered fields. Empirical observation of this peculiar yet seminal phenomenon was first reported by Wood \cite{Wood1902,Wood1935}
and explained decades later as forced resonances via comprehensive FB frameworks \cite{Fano1941,Hessel1965}. This concept inspired a massive investigation of periodic and quasi-periodic interfaces, laying the foundations for modern applications of electromagnetic-field manipulation, such as aperture synthesis of LW and holographic metasurface antennas \cite{Fong2010,Patel2011,Minatti2011,Morote2014,Monticone2015,Minatti2016Flat,Minatti2016Synthesis,Smith2017,Sanchez2018,Xu2023}. Recently, a nonlocal extension of Wood's anomaly beyond regular boundary conditions has been reported as well \cite{Shaham2021}.

Versatile and useful as it is, the above traditional scheme of crafting static unbiased media generally entails inherent limitations, e.g., reciprocity \cite{Collin2001,Caloz2018,Asadchy2020}, bandwidth bounds \cite{Foster1924,Bode1945,Chu1948,Fano1950,Harrington1960,Rozanov2000}, and lack of gain. One extremely resourceful paradigm to relax such constraints features modulating the constitutive properties of materials in time \cite{Galiffi2022}, typically by applying external bias. Notably, it provides one with the means to break time-translation or time-reversal symmetries, corresponding, respectively, to energy (frequency) conservation and reciprocity \cite{Collin2001,Yu2009,Hadad2015,Hadad2016,Sounas2017,Caloz2018,Cardin2020,Asadchy2020,Wang2020Theory,Wang2020Nonreciprocity,Hadad2024}.

Apart from transcending the performance of static media \cite{Yu2009,Hadad2015,Hadad2016,Sounas2017,Cardin2020,Wang2020Theory,Wang2020Nonreciprocity,Shlivinski2018,Hrabar2020,Firestein2022,Hayran2023,Hadad2024,Fritts2025,Ciabattoni2025}, time-varying composites introduce distinct characteristics of their own, as their ``microscopic'' scattering process is manifested through the exchange of frequency \cite{Morgenthaler1958,Mendonca2002,Galiffi2022}, rather than wavevector. Thereby, the abundant benefits offered by this novel degree of freedom have catalyzed the exploration of pivotal phenomena, both in theory and experiment, such as time reflection \cite{Morgenthaler1958,Mendonca2002,Moussa2023,Jones2024}---the temporal counterpart of Fresnel reflection---and momentum bandgaps \cite{Biancalana2007,Zurita2009,Martinez2016,Lustig2018,Dikopoltsev2022,Lyubarov2022,Wang2023}---the temporal analog of energy bandgaps in photonic crystals.

In light of such a clear duality between wavevector (space modulation) and frequency (time modulation) transitions, spatiotemporal \cite{Hadad2015} and temporal \cite{Galiffi2020} analogs of Wood's anomaly have been theoretically predicted as well. That is, an incident wave from the farfield can be strongly coupled to an SW across the light line and back rather through a frequency shift provided by a time-periodic grating than a momentum shift provided by a space-periodic grating (or in conjunction with it). Granted, temporal interfaces may surrogate their spatial counterparts for this task and thus alleviate common practical challenges of sensitivity to discretization and fabrication tolerances \cite{Galiffi2020}. Nevertheless, they come at the substantial cost of other realistic challenges typical of time-varying metamaterials.

Even in the radio-frequency (RF) and microwave bands, where size and loss tend to be relatively affordable, implementing a time grating in practice requires embedding a large number of time-modulated elements on a surface or in a bulk \cite{Cardin2020,Wu2020,Moussa2023,Wang2023,Park2022,Jones2024}. Feeding the modulation signal to this distributed cluster of elements naturally requires complex and power-hungry networks. A fortioiri, nontrivial effort must often be made during design to isolate these networks from the electromagnetic environment to avoid mutual interference \cite{Wu2020,Wang2023}. Such decoupling is mostly achieved by the use of ground planes, which hinder transmissive applications.

By now, Wood's anomaly of the spatial type can be undoubtedly viewed as the cornerstone of SW and LW applications \cite{Fong2010,Patel2011,Minatti2011,Morote2014,Monticone2015,Minatti2016Flat,Minatti2016Synthesis,Smith2017,Sanchez2018,Xu2023}; likewise, the temporal type \cite{Galiffi2020} can be inferred equally as instrumental for the analogous synthesis of ``temporal apertures'' for avant-garde wave manipulation. However, the true potency of this avenue, beyond the simple analogy to its spatial counterpart, is presently shrouded by several gaps that impede significant progress. From the fundamental perspective, no direct experimental observation of temporal Wood's anomaly and its intricate mechanism has been reported thus far. Moreover, while exhibiting interesting effects, e.g., asymmetric scattering \cite{Tsai2022}, the specific configurations studied so far have not yet channeled the enhanced intrinsic capabilities of time modulation, e.g., inherently tunable violation of energy conservation, in the context of temporal Wood's anomaly.

\begin{figure*}
    \includegraphics[width=\textwidth]{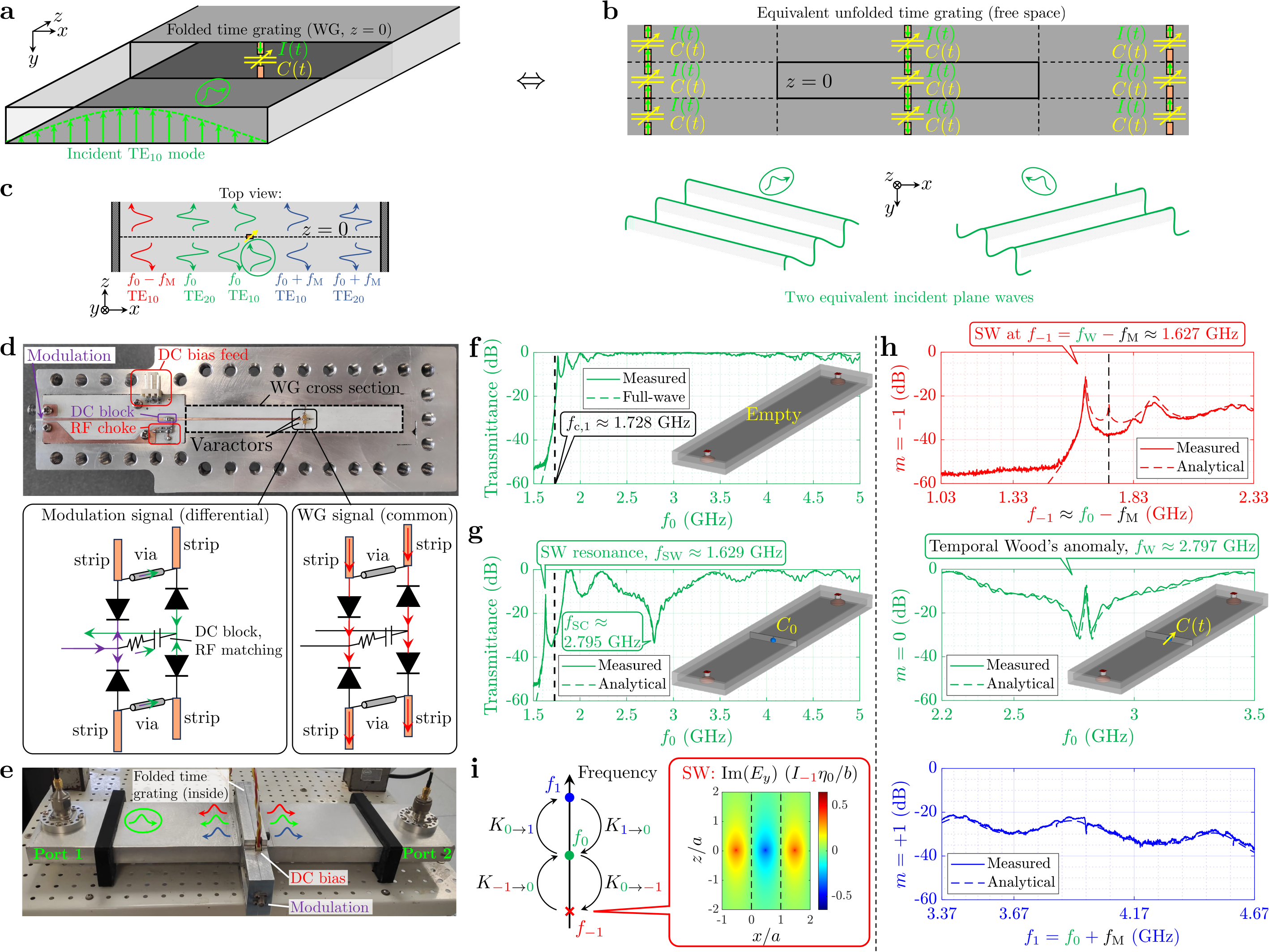}
\caption{\textbf{a} Folded time grating of a single time-modulated element $C(t)$ inside a WG ($0<x<a$, $0<y<b$). An incident TE$_{10}$ mode (green electric-field arrows) induces time-varying current $I(t)$. \textbf{b} Equivalent unfolded time grating of infinitely many time-modulated elements in free space ($-\infty<x<\infty$, $-\infty<y<\infty$): image currents and two obliquely incident plane waves. \textbf{c} Conceptual scattering schematic: incident TE$_{10}$ mode (green, encircled) and representative scattered TE$_{n0}$ modes of the $m=-1$ (red), $m=0$ (green), and $m=1$ (blue) time harmonics. \textbf{d} Realization of the time-modulated interface (front view): DC and modulation signal are fed differentially (left inset), whereas interaction with the WG fields is manifested by the common mode (right inset). \textbf{e} Experimental setup: coaxial excitation of frequency $f_0$ in Port 1 is converted to an incident TE$_{10}$ wave (green, encircled) and scattered from the folded time grating as in \textbf{c}. The transmitted TE$_{10}$ modes of all frequencies are collected by Port 2. \textbf{f} Reference measurement of voltage transmittance (solid trace) through an empty WG, compared to simulated results (dashed trace) of the CST WG and adapter model (depicted); the black dashed line marks the fundamental cutoff frequency (also in \textbf{g} and \textbf{h}). \textbf{g} Transmittance measurement (solid) with static capacitance ($6$ V reverse DC bias per varactor), compared to analytical predictions (dashed). Surface-wave and SC resonances are observed.  \textbf{h} Measured (solid) and analytical (dashed) transmittance for the $m=-1$ (red), $m=0$ (green), and $m=1$ (blue) time harmonics formed by sinusoidally modulating the capacitance in time ($f_{\mathrm{M}}=1.17$ GHz). Temporal Wood's anomaly appears as a transmissive peak at the Wood frequency $f_{\mathrm{W}}\approx 2.797$ GHz in the fundamental time harmonic (green), accompanied by strong coupling to the SW in the $m=-1$ harmonic (red) at $f_{-1}=f_{\mathrm{W}}-f_{\mathrm{M}}\approx 1.627$ GHz ($\approx f_{\mathrm{SW}}$). \textbf{i} Coupling process schematic: the time harmonics are consecutively coupled through the primitive coupling coefficients $K_{m' \to m}$. Temporal Wood's anomaly occurs when the downconverted frequency $f_{-1}$ resides near the SW pole ($f_{\mathrm{SW}}$, red ``$\times$''). Inset: electric field $\mathrm{Im}(E_{y})$ versus $x$ and $z$ of the SW at $f_{-1}=1.627$ GHz ($I_{-1}$ is the current at frequency $f_{-1}$); black dashed lines at $x=0,a$ mark the locations of the WG walls.
}
\label{Fig:Fig1}
\end{figure*}

In this paper, we fill these gaps and reveal novel elemental aspects of temporal Wood's anomaly by proposing, analyzing, and measuring a \emph{compact} setup: we enclose a printed copper strip loaded by a time-varying capacitor inside a metallic waveguide (WG) and, so, employ electromagnetic image theory \cite{Killamsetty2021} to effectively span an entire time-modulated planar interface from a \emph{single} time-modulated element (Figs.\ \ref{Fig:Fig1}a and b). By this, we drastically reduce complexity and power consumption, vacating our attention to decoupling the modulation circuitry from the electromagnetic WG environment and facilitating wideband transmissive operation.

After establishing a rigorous FB scattering analysis and setting up our experimental apparatus, we first identify the static SW resonance of the loaded-strip interface, through which temporal Wood's anomaly is to be channeled. Subsequently, we modulate the load capacitance in time and observe the temporal Wood phenomenon experimentally, including the emergence of the down-converted SW as the main part of this process. Noting excellent agreement between our results and detailed theory, we deepen the analysis to reveal and explain their intricate characteristics.

In addition to this direct experimental validation of the theoretical concept proposed in \cite{Galiffi2020},
we carefully harness our insightful analytical formulation to contrive a new type of temporal Wood's anomaly, where coupling to the negative-frequency branch \cite{Fritts2025,Feinberg2025} of the SW manifests tunable, negative-resistance parametric amplification \cite{Collin2001}. While temporal Wood's anomaly in its standard form of positive SW frequencies \cite{Galiffi2020,Tsai2022} allows one to surrogate spatial gratings of passive applications, the previously unexplored negative-frequency regime found herein grants the possibility to integrate an inherent benefit of time modulation---violation of energy conservation that yields gain---which has not been accessible in such past endeavors. Overall, beside the experimental milestone in this area, our results pave the path to enhanced dynamic spectral shaping of ``temporal apertures''---the analogs of synthesized spatial apertures \cite{Fong2010,Patel2011,Minatti2011,Morote2014,Monticone2015,Minatti2016Flat,Minatti2016Synthesis,Smith2017,Sanchez2018,Xu2023}---which may prove beneficial to frequency-agile filtering and LW antennas.

\section*{Results}
\label{Sec:Results}

\subsection*{Analysis}
\label{Subsec:Analysis}
We consider an infinitely long ($-\infty <z<\infty$) rectangular WG of perfect-electric-conductor (PEC) walls of width $a$ ($0<x<a$) and height $b$ ($0<y<b$, $a>b$), as depicted in Fig.\ \ref{Fig:Fig1}a. For our purposes herein, we focus on a standard $y$-invariant and $y$-polarized transverse-electric (TE) configuration ($\partial_y=0$ and $E_{x}=E_{z}=H_{y}\equiv 0$), in which the fields can be completely spanned by the TE$_{n0}$ modes of the WG \cite{Collin2001}; all other modes are suppressed.

To implement the folded time grating, we center a $y$-oriented, thin metallic strip of width $w$ on the $z=0$ plane inside the WG, at $x=x_{\mathrm{s}}=a/2$ (Fig.\ \ref{Fig:Fig1}a); the strip is loaded by a lumped time-varying capacitor of capacitance $C(t)$. Specifically, we apply periodic forms of time modulation, which may generally be expressed via a Fourier series \cite{Wu2020,Galiffi2022,Wang2020Theory,Wang2023},
\begin{equation}
\label{Eq:C(t)}
    C(t)=\sum_{l=-\infty}^{\infty}C_{l}e^{jl\Omega t},
\end{equation}
where $\Omega=2\pi f_{\mathrm{M}}$ is the (angular) frequency of the modulation, and
\begin{equation}
\label{Eq:C_m}
    C_{l}=\frac{\Omega}{2\pi}\int_{0}^{2\pi/\Omega}C(t)e^{-jl\Omega t}dt,\quad l\in\mathbb{Z}
\end{equation}
are the associated complex Fourier amplitudes, which satisfy $C_{-l}=C_{l}^{*}$, such that the capacitance $C(t)$ is real valued at all times; $(\cdot)^{*}$ denotes complex conjugation.

Next, we launch a TE$_{10}$ mode of frequency $\omega_0=2\pi f_0$ from $z<0$ toward the loaded strip (Fig.\ \ref{Fig:Fig1}a). The $y$-polarized electric field of this incident wave is expressed via \cite{Collin2001}
\begin{equation}
\label{Eq:Ey_inc}
    E_{y}^{\mathrm{inc}}\left(\vec{r};t\right)=E_{01}^{\mathrm{inc}}\sin\left(\frac{\pi}{a}x\right)e^{-j\beta_{01}z}e^{j\omega_0 t},
\end{equation}
where $E_{01}^{\mathrm{inc}}$ is the amplitude at $(x,z)=(a/2,0)$,
\begin{equation}
\label{Eq:beta_01}
    \beta_{01}=\sqrt{\left(\frac{\omega_0}{c}\right)^{2}-\left(\frac{\pi}{a}\right)^{2}}
\end{equation}
is the propagation constant associated with this TE$_{10}$ mode at frequency $\omega_0$, and $c$ is the speed of light. The impinging field of Eq.\ (\ref{Eq:Ey_inc}) induces time-dependent current $I(t)$ in the strip (Fig.\  \ref{Fig:Fig1}a), which, in turn, emanates secondary (scattered) fields inside the WG.

Before proceeding further, it would be instructive to elucidate the concept of a \emph{folded} time grating and the motivation behind it. As illustrated in Fig.\ \ref{Fig:Fig1}b, fictitious image currents can be placed outside the WG to nullify the tangential field exerted by the real current on the wall between them (PEC boundary condition); in turn, each such image current induces an additional, farther image current to nullify the tangential field on the opposite WG wall as well. This step can be infinitely repeated, such that an entire spatially periodic grating is effectively reproduced \cite{Killamsetty2021}. Additionally, the $\sin\left(\frac{\pi}{a} x\right)e^{-j\beta_{01}z}$ term in Eq.\ (\ref{Eq:Ey_inc}) features a standing wave along $x$ in the incident TE\textsubscript{10} mode (Fig.\ \ref{Fig:Fig1}a); it can be interpreted as the interference of two plane waves (Fig.\ \ref{Fig:Fig1}b), $\frac{1}{2j}\left[e^{-j(-\frac{\pi}{a}x+\beta_{01}z)}-e^{-j(\frac{\pi}{a}x+\beta_{01}z)}\right]$, which propagate forward at opposing oblique angles, so that the node planes of the interference coincide with the vertical WG walls.

Once the periodic configuration of Fig.\ \ref{Fig:Fig1}b is established, the WG walls may be removed and the domain of the $x$ and $y$ coordinates can be extended from the interior of the WG to the entire $xy$-plane. This forms an equivalent \emph{unfolded} electromagnetic setup of an entire metagrating plane \cite{Killamsetty2021}, situated in free space, with infinitely many time-varying loads. In fact, Fig.\ \ref{Fig:Fig1}b represents the prevalent paradigm of realizing time-varying media by embedding many time-modulated elements on a plane (or in a bulk). However, as discussed in the Introduction, the folded WG configuration in Fig.\ \ref{Fig:Fig1}a offers a compact and economical platform to demonstrate such effects, as it requires only a single time-modulated element to achieve identical field distribution and scattering. With this motivation in mind, let us resume our scattering analysis.

Since the height of the WG $b$ is set deeply subwavelength, we may apply the homogenization approximation \cite{Killamsetty2021,Hadad2024} and assume that the current distribution along the strip is $y$-independent, $I(y;t)\equiv I(t)$, in line with the global TE configuration. Furthermore, owing to the time periodicity of the modulation, we may express this current waveform as a superposition of FB time harmonics \cite{Galiffi2020},
\begin{equation}
\label{Eq:I}
    I(t)=\sum_{m=-\infty}^{\infty}I_{m}e^{j\omega_m t},
\end{equation}
where
\begin{equation}
\label{Eq:omega_m}
    \omega_m=2\pi f_{m}=\omega_0+m\Omega,\quad m\in\mathbb{Z}
\end{equation}
and $I_m$ are the frequency and (yet-unknown) amplitude associated with the $m$th FB time harmonic, respectively.

Following the standard thin-wire approximation \cite{Tretyakov2003,Killamsetty2021,Hadad2024}, the current flowing through the strip effectively forms a line source concentrated at $(x,z)=(x_{\mathrm{s}},0)$ (Dirac impulse distribution along $x$ and $z$). This current density can be represented as a Fourier series of sinusoidal space harmonics, corresponding to the fundamental and higher-order TE$_{n0}$ modes of the WG ($n\in\mathbb{N} \backslash\{0\}$). Hence, an expression for the secondary fields emanating from the strip can be readily found through standard mode matching \cite{Collin2001,Killamsetty2021} (Supplementary Note A.1). For a centered strip of $x_{\mathrm{s}}=a/2$ only odd WG modes ($n=1,3,5,...$) arise due to symmetry,
\begin{equation}
\label{Eq:Ey_scat}
    \begin{aligned}
        E_{y}^{\mathrm{scat}}(\vec{r};t)=\sum_{m=-\infty}^{\infty}&\sum_{n=1,3,...}\left\{ -\frac{\eta_{0}\omega_{m}I_{m}}{ac\beta_{mn}}(-1)^{(n-1)/2}\right.\\
        &\left. \times \sin\left(\frac{\pi n}{a}x\right)e^{j\omega_m t}e^{-j\beta_{mn}|z|}\right\},
    \end{aligned}
\end{equation}
where
\begin{equation}
\label{Eq:beta_mn}
    \beta_{mn}=\sqrt{\left(\frac{\omega_m}{c}\right)^{2}-\left(\frac{\pi n}{a}\right)^{2}}
\end{equation}
is the propagation factor associated with the TE$_{n0}$ mode ($n\in\mathbb{N}\setminus\left\{0\right\}$) scattered by this $m$th time harmonic. Conceptual scattering schematic is provided in Fig.\ \ref{Fig:Fig1}c.

Each of the TE$_{n0}$ WG modes is associated with a cutoff frequency, $\omega_{\mathrm{c},n}=\frac{\pi n}{a}$. If, for some $m$ and $n$, the $m$th time harmonic is of frequency greater than the cutoff frequency of the TE$_{n0}$ WG mode, i.e., $\omega_m>\omega_{\mathrm{c},n}$, then this space-time harmonic assumes a real-valued propagation constant $\beta_{mn}$ allowing farfield radiation; otherwise, the propagation constant is purely imaginary, resulting in an exponentially decaying evanescent wave. In the unfolded free-space setup (Fig.\ \ref{Fig:Fig1}b), the TE$_{n0}$ WG mode corresponds to the $n$th spatial diffraction order, so that evanescent harmonics of large index $n$ are confined to the vicinity of the grating and progress along the $\pm x$-directions as SWs. The cutoff frequency is therefore equivalent to grazing free-space progression that draws the border (light line) between propagating and evanescent waves. It is by coupling to such surface waves and back, through frequency (wavevector) exchange, that the temporal (spatial) Wood's anomaly emerges.

Following Eq.\ (\ref{Eq:Ey_scat}), one obtains a comprehensive solution for the electromagnetic problem upon finding the values of the current amplitudes $I_m$ ($m\in\mathbb{Z}$). Namely, the reflection coefficient into the $mn$th space-time harmonic scattered back toward $z<0$ can be obtained via
\begin{equation}
\label{Eq:Gamma}
    \Gamma_{mn}=-\frac{\eta_0\omega_{m}I_{m}}{ac\beta_{mn}E_{01}^{\mathrm{inc}}}(-1)^{(n-1)/2}\quad \text{(odd $n$)},
\end{equation}
and the forward transmission coefficient toward $z>0$ may be found through
\begin{equation}
\label{Eq:tau}
    \tau_{mn}=\delta_{m0}\delta_{n1}+\Gamma_{mn},
\end{equation}
where $\delta_{ij}$ is the Kronecker delta. To obtain the current amplitudes $I_m$, we follow the constitutive relation \cite{Collin2001,Wu2020,Wang2020Theory,Hadad2024}
\begin{equation}
\label{Eq:Constitutive}
    \begin{aligned}
        I(t)=\frac{d}{dt}\left\{bC(t)\left[E_y^{\mathrm{inc}}\left(x=x_{\mathrm{s}},z=0;t\right)\right.\right. \\
        +\left.\left.E_y^{\mathrm{scat}}\left(x=x_{\mathrm{s}},z=r_{\mathrm{eff}};t\right)\right]\right\},
    \end{aligned}
\end{equation}
which states that the instantaneous current equals the time derivative of the instantaneous charge stored in the capacitor, where $r_{\mathrm{eff}}=w/4$ is the effective radius associated with the thin strip \cite{Tretyakov2003,Barkeshli2004,Killamsetty2021,Hadad2024}.

Next, we substitute Eqs.\ (\ref{Eq:C(t)}), (\ref{Eq:Ey_inc}), and (\ref{Eq:I})--(\ref{Eq:Ey_scat}) in Eq.\ (\ref{Eq:Constitutive}) (Supplementary Note A.2) and obtain an infinite set of equations for the unknown current harmonics $I_{m}$, $m\in\mathbb{Z}$; the equations are indexed by $l\in\mathbb{Z}$ and given via
\begin{equation}
\label{Eq:DiffEqns}
    \sum_{m=-\infty}^{\infty}\left(\delta_{lm}-\omega_{l} C_{l-m}\omega_{m}L_{m}\right)I_{m}=j\omega_{l}C_{l}bE_{01}^{\mathrm{inc}},
\end{equation}
where we interpret
\begin{equation}
\label{Eq:L_m}
    L_{m}\left(\omega_{m}\right)=\frac{\mu_0b}{a}\sum_{n=1,3,...}\frac{1}{j\beta_{mn}}e^{-j\beta_{mn}r_{\mathrm{eff}}}
\end{equation}
as the equivalent (complex-valued and $\omega_{m}$-dependent) self inductance of the strip at the frequency $\omega_m$ in the presence of the WG walls.


For each given set of setup parameters ($a$, $b$, $w$, $x_{\mathrm{s}}=a/2$, $f_0$, $f_{\mathrm{M}}$, and $\{ C_l\}_{l\in\mathbb{Z}}$), we may formulate Eq.\ (\ref{Eq:DiffEqns}) in matrix form and numerically apply truncation and inversion to obtain the current amplitudes $I_m$, which subsequently yield the scattering coefficients via Eqs.\ (\ref{Eq:Gamma}) and (\ref{Eq:tau}). Having achieved this rigorous analytical formulation to evaluate the scattering properties of our device, we now proceed to introduce the experimental setup to observe temporal Wood's anomaly and reveal its extended features.

\subsection*{Design and experiment}
\label{Subsec:Design}
To set the stage for comparison between the analytical and experimental results, we briefly outline the essentials of the fabricated apparatus, measurement procedure, and data processing; more details are provided in Methods and Supplementary Notes D and E. We construct a rectangular aluminum WG based on a standard reduced-height WR-340 model (Figs.\ \ref{Fig:Fig1}d and e) of $a=86.36$ mm width and $b=10.40$ mm height. The folded time-modulated interface at $z=0$ (Fig.\ \ref{Fig:Fig1}a) is implemented via a balanced bridge of four commercial varactor diodes mounted on two copper strips of width $w=0.635$ mm printed on each side of a thin commercial printed-circuit-board (PCB) substrate (Fig.\ \ref{Fig:Fig1}d).

This bridge formation enables excellent decoupling between the modulation network and the electromagnetic environment, as desired for transmissive operation: reverse direct-current (DC) bias and RF modulation are applied through the differential (antisymmetric) mode supported by the bridge (left inset of Fig.\ \ref{Fig:Fig1}d), whereas the interaction of the loaded strip with the WG fields is carried through the common (symmetric) mode (right inset). Namely, the differential mode features two opposing and closely spaced current paths, such that the total interference due to the presence of the modulation network is negligible. This also allows us to match the modulation network in a wide frequency range without undesirably loading the common-mode capacitance and eroding its effect.

The incident wave is launched from Port 1 (Fig.\ \ref{Fig:Fig1}e), in which a suitable in-house designed adapter effectively converts an input coaxial voltage signal into a TE$_{10}$ mode (Supplementary Note D.1). Complementarily, the transmitted TE$_{10}$ waves of relevant FB frequencies $\omega_m$ are simultaneously collected by Port 2 (Fig.\ \ref{Fig:Fig1}e) and converted, by another coaxial adapter, to a voltage signal fed to a spectrum analyzer for detection.

As a reference for interpreting and analyzing the next results, the green solid trace in Figure \ref{Fig:Fig1}f shows the measured voltage transmittance from Port 1 to Port 2 versus frequency, in the range of $1.5$--$5$ GHz, through an empty WG, i.e., when the loaded-strip circuit is entirely removed. Based on this measurement (and others, Supplementary Note D.1), we establish an accurate full-wave model of the WG and its coaxial adapters (CST Studio Suite, depicted in Fig. \ref{Fig:Fig1}f), which yields the simulated results shown as a green dashed trace in Fig.\ \ref{Fig:Fig1}f, practically coinciding with the measured solid trace.

The WG width in this simulation (and in all the theoretical calculations to follow) that yields the best concurrence with the above experimental measurements is $a=86.77$ mm (extremely close to a caliper measurement of ${\sim 86.36}$ mm, ${\sim 0.5\%}$ error). This results in a fundamental cutoff frequency of $f_{\mathrm{c},1}=\frac{\omega_{\mathrm{c},1}}{2\pi}=\frac{c}{2a}\approx 1.728$ GHz (dashed black line). Below cutoff, the transmittance decreases exponentially when the frequency is lowered due to increase in the attenuation constant of the fundamental TE$_{10}$ mode [Eq.\ (\ref{Eq:beta_01})]. Above cutoff, nearly total transmission is observed, slightly deterred by small inevitable loss and minor ripples due to parasitic reflections off the coaxial adapters.

To provide accurate comparison between theory and measurement, also for the experiments to follow, we incorporate WG and adapter effects (propagation phase/attenuation, conversion from fields to voltage and vice versa, and adapter reflections) via our reliable full-wave electromagnetic model into the predictions of the Analysis section (Supplementary Note D.2). Importantly, note that the effects due to the time-modulated $z=0$ interface are evaluated in a completely analytical manner (FB framework above); the parasitic practical setup effects are only extracted from the full-wave simulation and then applied to the original analytical results to yield accurate interpretation of measurements. Also note that the ports are located far enough from the time-modulated interface, such that the dominant WG mode that reaches from one to the other is TE$_{10}$, which justifies the pure TE$_{10}$ excitation assumption in our analytical formulation (Supplementary Note D.2.1). This means that the ratio between the voltage detected at frequency $\omega_m$ in Port 2 to the applied voltage of frequency $\omega_0$ at Port 1 represents the TE$_{10}$ transmission coefficients $\tau_{m1}$ in Eq.\ (\ref{Eq:tau}).

Furthermore, since our analytical model does not consider the presence of the (ultrathin) PCB substrate, we effectively incorporate its thickness, $t_{\mathrm{PCB}}=0.508$ mm, by modifying the effective wire radius in Eqs.\ (\ref{Eq:Constitutive}) and (\ref{Eq:L_m}) from $r_{\mathrm{eff}}=w/4$ to $r_{\mathrm{eff}}=t_{\mathrm{PCB}}/2+w/4$ in our calculations to follow (Supplementary Note E.2). We are now prepared to reinstate the loaded-strip interface and investigate the resulting phenomena of interest.

\subsection*{Static load capacitance: SW resonance}
\label{Subsec:Static}
To lay the foundations for the appearance of temporal Wood's anomaly in our setup, let us first study the degenerate static case of time-invariant load capacitance, $C(t)\equiv C_0$, i.e., $C_{l}=C_0\delta_{l0}$ (see Eq.\ (\ref{Eq:C(t)})). In this scenario, no higher-order FB time harmonics emerge, such that Eq.\ (\ref{Eq:DiffEqns}) reduces to a single scalar equation, whose solution is
\begin{equation}
\label{Eq:I0_static}
    I_0=\frac{j\omega_0C_0bE_{01}^{\mathrm{inc}}}{1-\omega_0^2L_0(\omega_0)C_0}
\end{equation}
and, following Eqs.\ (\ref{Eq:Gamma}) and (\ref{Eq:tau}),
\begin{equation}
\label{Eq:Gamma0n_static}
    \Gamma_{01}=-\frac{j\eta_0\omega_0^2C_0b}{ac\beta_{01}\left[1-\omega_0^2L_0(\omega_0)C_0\right]},\quad \tau_{01}=1+\Gamma_{01}.
\end{equation}

The denominator term $1-\omega_0^2L_0(\omega_0)C_0$ introduces poles that correspond to (generally complex-valued) static resonant frequencies. Each of these free resonances is associated with a distinct eigenmode, self-sustained by the $z=0$ loaded-strip interface without external forces. One such eigenmode of central importance corresponds to a purely reactive nearfield wave confined to the loaded-strip interface---the equivalent of a laterally standing SW in the unfolded grating (see inset of Fig.\ \ref{Fig:Fig1}i)---of a \emph{real-valued} resonant frequency $\omega_{\mathrm{SW}}=2\pi f_{\mathrm{SW}}$ below cutoff, $0<\omega_{\mathrm{SW}}<\omega_{\mathrm{c},1}$. Its existence and uniqueness in this frequency range can be easily proven (Supplementary Note B.2).
In reality, the presence of small dissipation slightly shifts this resonant frequency outside the real frequency line, following a small imaginary appendage that corresponds to exponential decay in time due to loss.

To observe and characterize this pole in our device, we set the reverse DC bias of the varactor diodes to $6$ V each and measure the voltage transmittance from Port $1$ to Port $2$, as before, in the range of $1.5$--$5$ GHz (Fig.\ \ref{Fig:Fig1}g, green solid trace). Below the cutoff frequency (1.728 GHz, black dashed line), we identify a sharp resonant peak ($\approx-14.5$ dB) appearing at $f_{\mathrm{SW}}\approx 1.629$ GHz on top of the previously measured exponential roll-off in the empty WG.

Above cutoff, we note total transmission (up to small parasitic ripples due to reflection off the coaxial adapters) followed by another resonant drop around $f_{\mathrm{SC}}\approx 2.795$ GHz. At this frequency, a different type of resonance is manifested (Supplementary Note B.3) as the reactive part of the strip inductance, $L_0'(\omega_0)=\Re[L_0(\omega_{0})]$, is perfectly canceled by the static capacitance $C_0$ in series. Note that, in contrast to the behavior at $f_{\mathrm{SW}}$, this resonance at $f_{\mathrm{SC}}$ is not an eigenmode of the system: the inductance includes an additional imaginary part above cutoff, $L_0''(\omega_0)=-\Im[L_0(\omega_0)]$ due to radiation (and dissipation) \cite{Tretyakov2003}, thus preventing the formation of a pole at this real-valued frequency. This resonance rather describes (nearly) perfect destructive interference between the incident and reflected TE$_{10}$ waves ($\Gamma_{01}\approx-1$), which yields vanishing transmittance (a zero instead of a pole). This resembles the behavior of a transmission line terminated by a serial LC load \cite{Pozar2012}: when the load resonates, an effective short-circuit (SC) condition ensues, leading to zero total voltage due to perfect destructive interference.

In comparison to these measured results, the theoretical prediction of transmittance with effective capacitance $C_0=510$ fF is plotted versus frequency as a green dashed trace in Fig.\ \ref{Fig:Fig1}g, exhibiting excellent agreement; to also account for dissipation, we followed a simple extension of the analysis (Supplementary Note C.4) featuring effective serial resistance of $R=1.4$ Ohm. The above effective values of static capacitance and resistance reasonably match the varactor diode specifications for the applied voltage (Supplementary Note E.2). In general, the SW and SC resonant frequencies can be tuned (albeit not independently) by varying the DC bias. With these essential insights at hand, we move on to demonstrate temporal Wood's anomaly and its extensions.

\subsection*{Temporal Wood's anomaly via sinusoidal time modulation}
\label{Subsec:Wood}
After successfully analyzing and experimentally validating the static case-study, we next consider sinusoidal time modulation,
\begin{equation}
\label{Eq:C_SinMod}
    \begin{aligned}
        C(t)&=C_0+C_1e^{j\Omega t}+C_{1}^{*}e^{-j\Omega t}\\
        &=C_0\left[1+2M\cos(\Omega t+\angle C_1)\right],
    \end{aligned}
\end{equation}
where $C_0$ is the DC offset, $C_1=|C_{1}|e^{j\angle C_{1}}$ is the fundamental complex amplitude according to the Fourier representation in Eq.\ (\ref{Eq:C(t)}) ($C_{-1}=C_{1}^{*}$), and $M=|C_{1}|/C_0$ is the modulation depth, limited in our setup to $0\leq M<0.5$ to maintain positive capacitance at all times. In this scenario, coupling to all higher-order time harmonics must be considered to obtain an exact solution, which invokes involved matrix inversion of infinitely many entries to solve the set of equations in Eq.\ (\ref{Eq:DiffEqns}). However, for our purposes of observing and explaining the appearance of temporal Wood's anomaly, we may consider coupling only between the dominant \cite{Galiffi2020,Hadad2024} $m=0,\pm 1$ time harmonics and neglect higher-order effects (verified below).



Before presenting the elaborated analysis, let us first recognize the phenomena we wish to capture by exploring a typical case study. We retain the previous DC bias of $6$ V and apply an additional sinusoidal tone of $f_{\mathrm{M}}=1.17$ GHz modulation frequency and $M\approx 4.7\%$ modulation depth (modulation voltage signal of 18 dBm, Supplementary Note E.2.2). We plot the measured (solid traces) and calculated (dashed traces) transmittance results for the $m=-1$ (red), $m=0$ (green), and $m=1$ (blue) time harmonics versus their respective frequencies, $f_m=f_0+mf_{\mathrm{M}}$ in Fig.\ \ref{Fig:Fig1}h when the frequency of the incident wave $f_0$ is swept above cutoff, along $2.2$--$3.5$ GHz.

For the fundamental $m=0$ time harmonic (green), we identify a new transmissive peak ($\approx -8.27$ dB) appearing at what we identify as the Wood frequency $f_{\mathrm{W}}\approx 2.797$ GHz, on top of the previous SC resonant drop (cf.\ Fig.\ \ref{Fig:Fig1}g). This peak occurs at the excitation frequency $f_0=f_{\mathrm{W}}$ for which the down-converted frequency of the $m=-1$ harmonic approximately meets the (positive) SW resonant frequency $f_{\mathrm{SW}}$, i.e., $f_{-1}=f_{\mathrm{W}}-f_{\mathrm{M}}\approx f_{\mathrm{SW}}$. This peak phenomenon is the \emph{embodiment of temporal Wood's anomaly} \cite{Hadad2015,Galiffi2020}: the back-and-forth coupling to the $m=-1$ time harmonic near its SW resonance below cutoff ($f_{\mathrm{SW}}$) imprints abrupt fluctuations in the intensity of the fundamental scattering, analogously to the spatial phenomenon \cite{Wood1902,Wood1935,Fano1941,Hessel1965}.

We rigorously confirm this observation by examining the transmitted $m=-1$ time harmonic (red plot, Fig.\ \ref{Fig:Fig1}h), measured at frequency $f_{-1}=f_0-f_{\mathrm{M}}$ during the same sweep of incident frequency $f_0$. We discern a clear sharp peak forming exactly at the SW resonant frequency $f_{-1}=f_{\mathrm{SW}}$. This is, in fact, a \emph{direct measurement} of the evanescent tail of the SW confined to the loaded-strip interface, which is coupled through frequency down-conversion enabled by the time-modulation \cite{Galiffi2020} (see also Fig.\ \ref{Fig:Fig1}i). It reaches Port 2 with noticeable amplitude ($\approx-14.7$ dB), despite suffering considerable WG attenuation at its frequency (Fig.\ \ref{Fig:Fig1}f). To further demonstrate the dominance of the coupling to the $m=-1$ time harmonic in this process, we show, for comparison, the transmittance to the $m=+1$ time harmonic (blue plot), measured at $f_{1}=f_0+f_{\mathrm{M}}$ for the same sweep. In the vicinity of temporal Wood's anomaly, the $m=1$ transmission remains below $-22$ dB, compared to the peak attained by the $m=-1$ harmonic (red plot). Considering that the $m=1$ time harmonic, of a propagating frequency above cutoff, does not suffer the severe WG attenuation imposed on its evanescent $m=-1$ counterpart, we deduce the dominance of the latter in the process.

All the experimental results exhibit excellent agreement with their respective theoretical calculations (dashed lines, Fig.\ \ref{Fig:Fig1}h). The analytically calculated $m=-1$ time harmonic (dashed red trace) displays minor discrepancies around its cutoff frequency ($1.728$ GHz), possibly due to substrate-related effects, which are not considered in the effective single-strip model (see the ``Design and experiment'' section). Overall, this experiment provides a successful and accurate observation of the previously predicted temporal Wood's anomaly in a folded time grating, including its underlying intricate SW mechanism, showing notable congruence with our theoretical analysis.

\begin{figure*}
    \includegraphics[width=\textwidth]{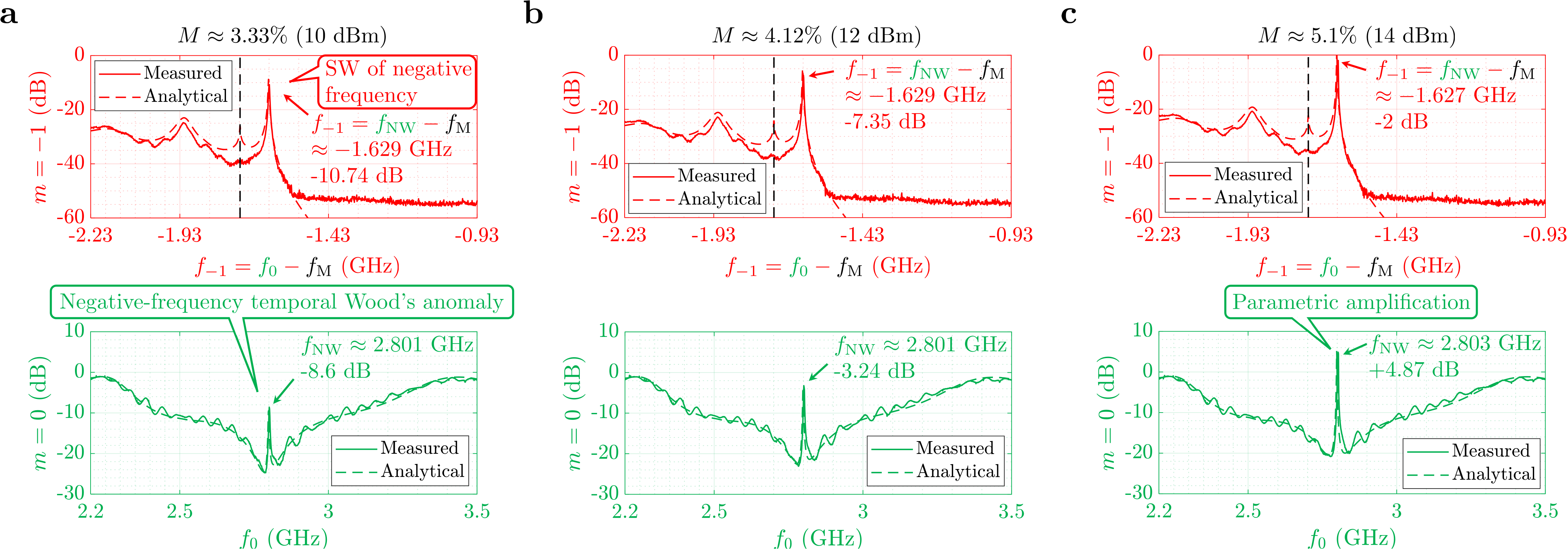}
\caption{Measured (solid traces) and analytical (dashed traces) voltage transmittance to the $m=-1$ (red) and $m=0$ (green) time harmonics for $6$ V reverse DC bias (as in Fig.\ \ref{Fig:Fig1}), $f_{\mathrm{M}}=4.43$ GHz modulation frequency, and \textbf{a} $M\approx3.33\%$ (10 dBm modulation signal), \textbf{b} $M\approx 4.12\%$ (12 dBm), and \textbf{c} $M\approx 5.1\%$ (14 dBm) modulation depths. This modulation frequency allows coupling near negative resonant SW frequencies (marked by a red box in \textbf{a} and red arrows in \textbf{a--c} that indicate the peak transmittance values and their frequency of occurrence), which results in another type of temporal Wood's anomaly (marked by green box in \textbf{a} and green arrows in \textbf{a--c} that indicate the peak transmittance values and their frequencies). The peak transmittance grows rapidly with the modulation depth $M$ and, particularly, features parametric amplification in \textbf{c}.}
\label{Fig:Fig2}
\end{figure*}

\subsection*{Analysis and extension of temporal Wood's anomaly}
\label{Subsec:Extension}
Having carefully identified the main phenomenon of interest, we now return to analyze it in more detail to reveal new insights and extend its scope. Considering only the interaction between the $m=0,\pm 1$ time harmonics in Eq.\ (\ref{Eq:DiffEqns}), as verified above, we analytically find the amplitude of the fundamental current harmonic (Supplementary Note C.3),
\begin{equation}
\label{Eq:I0_TimeMod}    I_0\!=\!\frac{jbE_{01}^{\mathrm{inc}}}{\omega_0L_0(\omega_0)}\!\!\left(\!\frac{1+K_{0\to 0}}{1-K_{0\to -1}K_{-1\to 0}-K_{0\to 1}K_{1\to 0}}-1\!\right)\!,
\end{equation}
where
\begin{equation}
\label{Eq:K}
    K_{m'\to m}=\frac{\omega_{m}C_{m-m'}\omega_{m'}L_{m'}(\omega_{m'})}{1-\omega_m^{2}L_m(\omega_m)C_0},
\end{equation}
evaluated for $m'\in\{0,\pm 1\}$ and $m\in\{0,\pm 1\}$, defines a dimensionless primitive coupling coefficient from the $m'$th to the $m$th time harmonic. These primitive coupling processes can be fathomed as if consecutively occurring to and fro, e.g., from $m'=0$ to $m=-1$ ($K_{0\to -1}$) and back from $m'=-1$ to $m=0$ ($K_{-1\to 0}$), and so on, simultaneously interfering with one another (Fig.\ \ref{Fig:Fig1}i). This repeated interference constructs a (formal) geometric series that yields the denominator in Eq.\ (\ref{Eq:I0_TimeMod}). Note, by the way, the indifference of Eq.\ (\ref{Eq:I0_TimeMod}) to the modulation phase $\angle C_1$, since ${K_{m'\to m}K_{m\to m'}\propto C_{m'-m}C_{m-m'}=|C_{m-m'}|^{2}}$.

Intriguingly, by virtue of the denominator in Eq.\ (\ref{Eq:K}), we note that in each such operation of a $K_{m'\to m}$ coupling, the destination harmonic $m$ of frequency $\omega_m$ drives an underlying resonance of the effective strip inductance $L_{m}(\omega_{m})$ in series with the static capacitance $C_0$. The closer the destination frequency $\omega_m$ to resonance the stronger the impact of this harmonic on the overall scattering. We may readily follow this insight to explain the role and dominance of the $m=-1$ time harmonic observed near the Wood frequency: since the frequencies $\omega_{0}$ and $\omega_{1}$ are located far away from the SW resonance, compared to $\omega_{-1}\approx\omega_{\mathrm{SW}}$ (Fig.\ \ref{Fig:Fig1}i), we may neglect $K_{0\to 1}K_{1\to 0}$ compared to $K_{0\to -1}K_{-1\to 0}$ in the denominator ${1-K_{0\to -1}K_{-1\to 0}-K_{0\to 1}K_{1\to 0}}$ in Eq.\ (\ref{Eq:I0_TimeMod}). 

This approximated denominator, ${1-K_{0\to -1}K_{-1\to 0}}$, may be tuned to override the static scattering behavior due to the $K_{0\to 0}$ term, especially for frequencies around the temporal Wood phenomenon, as observed above. To gain more insight about such possibilities, we may directly follow the definition of Eq.\ (\ref{Eq:K}), along with $C_{-1}=C_{1}^{*}$ and $M=|C_{1}|/C_0$, to find $K_{0\to -1}K_{-1\to 0}=M^{2}K_{0\to 0}K_{-1\to -1}$. Following this property and the aforementioned negligibility of $K_{0\to 1}K_{1\to 0}$, Eq.\ (\ref{Eq:I0_TimeMod}) may be approximated as
\begin{equation}
\label{Eq:I0_TimeMod_Rephrased}
    I_0\approx\frac{jbE_{01}^{\mathrm{inc}}}{\omega_0L_0(\omega_0)}\left(\frac{1+K_{0\to 0}}{1-M^{2}K_{0\to 0}K_{-1\to -1}}-1\right),
\end{equation}
thus introducing another pole at a different frequency (generally complex-valued) that satisfies
\begin{equation}
\label{Eq:Pole}
    M^2K_{0\to0}K_{-1\to -1}=1.
\end{equation}
In fact, analogously to control theory \cite{Franklin2010}, Eq.\ (\ref{Eq:I0_TimeMod_Rephrased}) can be interpreted as a closed-loop gain by an underlying feedback: the static $K_{0\to 0}$ coupling forms an open-loop gain, whereas $M^{2}K_{-1\to -1}$ manifests a feedback gain regulated by the parameter $M^2$. Equation (\ref{Eq:Pole}) is then a root-locus equation that describes the trajectories traced by the poles in the complex frequency plane as $M^{2}$ is swept.

For a given set of parameters, the complex frequency of this pole, $\omega_{\mathrm{p}}=\omega_{\mathrm{p}}'-j\omega_{\mathrm{p}}''$, corresponds to an eigenmode of $e^{\omega_{\mathrm{p}}''t}e^{j\omega_{\mathrm{p}}'t}$ time dependence. If $\omega_{\mathrm{p}}''<0$, this stable eigenmode decays in time, similarly to a damped harmonic oscillator. However, for $\omega_{\mathrm{p}}''=0$, parametric oscillation occurs at the real-valued frequency $\omega_{\mathrm{p}}'$ \cite{Collin2001} as the power supplied by the modulation signal precisely compensates the loss due to radiation and dissipation, similarly to the lasing condition. In this scenario, any external TE$_{10}$ excitation of nonvanishing amplitude $E_{01}^{\mathrm{inc}}$ at the resonant frequency $\omega_{\mathrm{p}}'$ would, in principle, force an infinite current $I_0$ [Eq.\ (\ref{Eq:I0_TimeMod_Rephrased})] and, consequently, infinite reflectance and transmittance; in practice, of course, saturation or breakdown of the varactors would prevent this. Nonetheless, slightly detuning the location of the pole toward the stable region enables parametric amplification that features a finite gain \cite{Collin2001}. The remaining class of eigenmodes, for which $\omega_{\mathrm{p}}''>0$, corresponds to unstable self-oscillations that grow exponentially in time, resembling the behavior inside a momentum bandgap \cite{Biancalana2007,Zurita2009,Martinez2016,Lustig2018,Dikopoltsev2022,Lyubarov2022,Galiffi2022,Wang2023}.

Following these insights, we may seek to exploit the violation of power conservation due to the time variance of the system and access parametric amplification by adequately tuning the temporal Wood's anomaly to extract power from the external modulation signal and radiate it in the fundamental time harmonic. To draw the original pole toward the real frequency line and facilitate gain, it is mandatory, in view of Eq.\ (\ref{Eq:Pole}), to allow the phase of $K_{-1\to -1}$ to compensate the phase of $K_{0\to 0}$ at resonance, i.e., to aspire $\angle K_{-1\to -1}+\angle K_{0\to 0}=0$ \cite{Franklin2010}. Due to the properties of Eqs.\ (\ref{Eq:L_m}) and (\ref{Eq:K}), it can be shown (Supplementary Note C.5) that this phase requirement can be approached for real-valued $\omega_{0}$ \emph{only} if $\omega_{0}$ and $\omega_{-1}$ are of \emph{opposite signs} and only if dissipation is present at $\omega_{-1}$.

In our convention where $\omega_0$ is positive, this newfound regime requires $\omega_{-1}$ to be set near the negative-frequency branch \cite{Fritts2025,Feinberg2025} of the SW resonance, $\omega_{-1}\approx-\omega_{\mathrm{SW}}$. This can be accomplished by matching the modulation frequency to satisfy $\Omega\approx\omega_{0}+\omega_{\mathrm{SW}}$, such that parametric amplification, whose gain may be tuned via the modulation depth $M$ [Eq.\ \ref{Eq:Pole})], ensues at $\omega_0$. Under the approximation of Eq.\ (\ref{Eq:I0_TimeMod_Rephrased}), such setup conditions can be shown equivalent to a virtual scenario where the SW coupling occurs at the positive branch of frequencies, $\omega_{-1}\approx +\omega_{\mathrm{SW}}$, albeit with negating the (dissipation) resistance therein (Supplementary Note C.5). In other words, coupling to the negative frequency of parametric gain emulates the time-reversed process of dissipation at the corresponding positive frequency. In fact, this notion of equivalent negative resistance is common in parametric amplifiers and oscillators \cite{Collin2001}, and it has recently been leveraged to increase the bandwidth of electrically small antennas \cite{Fritts2025}. Herein, however, it appears as a fundamental feature accessed by the collective planar resonance of the loaded strip with its images in the context of temporal Wood's anomaly.

We demonstrate this gain-inducing negative-frequency extension of temporal Wood's anomaly in Fig.\ \ref{Fig:Fig2} by increasing the modulation frequency used in the former configuration (Fig.\ \ref{Fig:Fig1}) to $f_{\mathrm{M}}=4.43$ GHz  and measuring the transmittance to the $m=0$ (green) and $m=-1$ (red) time harmonics versus the excitation frequency $f_0$, which is swept along 2.2--3.5 GHz. Figures\ \ref{Fig:Fig2}a--c show, respectively, the transmission measurements (solid traces, $m=0$ in blue and $m=-1$ red) for modulation depths of $M\approx 3.33\%$ (10 dBm modulation signal), $M\approx 4.12\%$ (12 dBm), and $M\approx 5.1\%$ (14 dBm). As expected, we recognize a resonant peak forming in the fundamental transmission (green) at the negative-Wood frequency, near $f_{\mathrm{NW}}\approx 2.802$ GHz, for which $f_{-1}=f_{\mathrm{NW}}-f_{\mathrm{M}}\approx-f_{\mathrm{SW}}$. This $m=0$ peak transmittance increases rapidly from $-10.74$ dB (Fig.\ \ref{Fig:Fig2}a) to $-3.24$ dB (Fig.\ \ref{Fig:Fig2}b) to +4.87 dB (Fig.\ \ref{Fig:Fig2}c, parametric amplification) as the modulation depth $M$ increases and drives the pole toward the real frequency line. The coupling to the $m=-1$ harmonic (red) follows a similar trend around the anticipated SW resonant frequency and rigorously captures the underlying mechanism that facilitates this negative-frequency Wood phenomenon with parametric amplification. The respective analytical predictions (dashed curves) excellently agree with measured results. These results complement our observation of the previously predicted temporal Wood's anomaly and confirm our comprehensive framework. They prove the viability of our folded WG configuration as a compact platform for demonstrating time-modulated effects, covering a wide range of scenarios, including new regimes of parametric amplification.



\section*{Discussion}
\label{Sec:Discussion}
To conclude, this work provides a direct experimental observation of temporal Wood's anomaly, thus confirming recent predictions of temporal SW-coupling processes analogous to classical momentum transitions through space gratings. Beyond this fundamental validation, we have uncovered a previously unexplored regime, in which coupling to the opposite SW frequencies yields tunable parametric amplification. In contrast to past configurations of spatial, and even temporal gratings of positive SW frequencies, this negative-frequency regime grants access to the inherent ability of time-varying systems to draw power from the modulation source and manifest gain. Supported by a rigorous, insightful, and accurate analytical framework, as well as a compact and economical WG-enclosed platform, our findings lay strong foundations for synthesizing ``temporal apertures,'' which may enable dynamic spectral shaping of frequency-agile filters and enhanced LW antennas.

\section*{Methods}
\subsection*{Analytical calculations}
The analytical calculations derived in the Analysis section and presented as dashed traces in Figs.\ \ref{Fig:Fig1}g, \ref{Fig:Fig1}h, and \ref{Fig:Fig2} were executed in Mathworks \textsc{Matlab}\xspace R2021a. The infinite summation in Eq.\ (\ref{Eq:L_m}) was truncated to a finite summation over $n=1,3,5,...,n_{\mathrm{max}}=501$, which was found to yield sufficient convergence. For the analysis of temporal Wood's anomaly in Fig.\ \ref{Fig:Fig1}h (dashed traces), only the $m=0,\pm 1$ time harmonics were considered. For the negative-frequency-SW temporal Wood's anomaly in Fig.\ \ref{Fig:Fig2} (dashed traces), only the $m=0,-1$ time harmonics were considered.

\subsection*{Experimental measurements}
The empty-WG measurement shown in Fig.\ \ref{Fig:Fig1}f was performed with the setup in Fig.\ \ref{Fig:Fig1}e, where Port 1 was coaxially fed by a $-10$ dBm time-harmonic signal produced by a commercial Keysight N5182B MXG Vector Signal Generator (9 kHz--6 GHz), whose frequency was swept in the range of $1.5$--$5$ GHz every $1$ MHz; the received signal in Port 2 was detected by a commercial Agilent Technologies N9000A CXA Signal Analyzer (SA) every $1$ MHz in the same frequency range of $1.5$--$5$ GHz, with resolution (RBW) and video (VBW) bandwidths of $\mathrm{RBW}=\mathrm{VBW}=1$ MHz (attenuation due to loss in the coaxial cable and an inner SA attenuator of 10 dB were calibrated out). We selected the ``max hold'' function of the SA, while sweeping the frequency provided by the signal generator. The dwell time for each frequency of the signal generator ($50$ ms) is set to be substantially longer than the sweep time of the SA ($5.32$ ms) to ensure its capturing in the measurement.

For the static-capacitance measurement shown in Fig.\ \ref{Fig:Fig1}g, we integrated the PCB interface with the varactor diodes into the WG, as in Fig.\ \ref{Fig:Fig1}d, and provided DC bias voltage by connecting a Hewlett Packard (HP) E3620A Dual Output DC Power Supply (that supports two 25 V, 1 A outputs) through a commercial Changjiang (CJT) connector of three pins that provided positive voltage ($+6$ V), DC ground ($0$ V), and negative voltage ($-6$ V) to the diode bridge (Figs.\ \ref{Fig:Fig1}d and e). The middle ground pin is connected through the WG chassis to the vias, such that all the varactor diodes were biased by the same reverse DC voltage ($6$ V). The signal excitation and detection through Ports 1 and 2 were performed similarly to the empty-WG measurements in the former paragraph.

For the temporal Wood's anomaly measurement in Fig.\ \ref{Fig:Fig1}h, we retained the 6 V DC bias and injected a $18$ dBm time-harmonic modulation signal of frequency $f_{\mathrm{M}}=1.17$ GHz through the coaxial modulation input (Figs.\ \ref{Fig:Fig1}d and e). The signal was provided by a Hittite Microwave Corporation HMC-T2100 10 MHz--20 GHz Signal Generator. The output RF power of the signal generator was set after calibrating out the loss due to the cable, such that the intended power of $18$ dBm reached the input modulation connector of the circuit. The signal generator and SA settings were configured similarly to the empty-WG setup in the beginning of this ``Experimental measurements'' subsection. Subject to the ``max hold'' configuration, the SA simultaneously captured the signals of all three frequencies, $\omega_{-1}$, $\omega_{0}$, and $\omega_{1}$, corresponding to the the $m=0,\pm1$ time harmonics. The data in these frequencies were later resolved to produce the red, green, and blue plots in Fig.\ \ref{Fig:Fig1}h. To avoid data overwriting between adjacent time harmonics, the $2.2$--$3.5$ GHz sweep was divided into two different sweeps of $2.2$--$2.5$ GHz and $2.5$--$3.5$ GHz (and later reunited to produce Fig.\ \ref{Fig:Fig1}h).

For the subsequent measurements of temporal Wood's anomaly with negative-frequency surface-wave coupling (Fig.\ \ref{Fig:Fig2}), the modulation frequency of the Hittite HMC-T2100 Signal Generator was modified to $f_{\mathrm{M}}=4.43$ GHz, and the amplitude was separately set to provide $10$ dBm (Fig.\ \ref{Fig:Fig2}a), $12$ dBm (Fig.\ \ref{Fig:Fig2}b) and $14$ dBm (Fig.\ \ref{Fig:Fig2}c) power in the coaxial input of the modulation signal (Fig.\ \ref{Fig:Fig1}d and e). The settings of the signal generator and SA for the excitation and detection of signals through Ports 1 and 2 were kept the same as in our previous measurements. This time, the signals detected by the SA were at frequencies $\omega_{0}$ of the fundamental $m=0$ time harmonic and $|\omega_{-1}|$, which represents the $m=-1$ time harmonic of the negative frequency $\omega_{-1}$ (as the fields and voltages are always real-valued functions of time and, hence, must contain also the complex conjugated signal of $e^{-j\omega_{-1}t}=e^{j|\omega_{-1}|t}$ time-dependence). No overlapping between the frequencies $\omega_0$ and $|\omega_{-1}|$ occurs in this frequency sweep, so the entire sweep was performed without separation to sub-sweeps.

\subsection*{Incorporation of WG and adapter effects}
To incorporate WG and adapter effects, as mentioned in the ``Design and experiment'' subsection (Results), we first measured all the scattering parameters ($S_{11}$, $S_{12}$, and $S_{21}=S_{12}$, \cite{Pozar2012}) between the coaxial Ports 1 and 2 of the empty WG (similarly to the SA measurement of Fig.\ \ref{Fig:Fig1}f) by connecting them to a commercial Agilent Technologies N5230A 10 MHz--40 GHz vector network analyzer (VNA, see Supplementary Note D.1). The VNA was set to measure these scattering properties in the $1.5$--$5$ GHz band; the frequency was sampled every $1$ MHz with resolution bandwidth of $\mathrm{RBW}=1$ kHz, and the excitation power in Port 1 was set to $-10$ dBm.

Next, based on these measurements, we constructed a full-wave model of the WG and its adapters in CST Studio Suite 2022 (SIMULIA, Dassault Systems), as also explained in the ``Design and experiment'' subsection and Supplementary Note D.2.1. The coaxial connectors were defined in CST as Waveguide ports of a single WG mode (the fundamental transverse electromagnetic, TEM, mode of 50 Ohm characteristic impedance). Based on this accurate full-wave model, we extracted the fields-to-voltage and voltage-to-field conversion factors of the WG adapters, as well as the reflection off them, from CST (Supplementary Note D.2.1) and analytically incorporated their effects on the scattering properties of the time-modulated loaded-strip interface, as elaborated in Supplementary Note D.2.2.

\section*{Acknowledgements}
A.\ S.\ acknowledges support from the Andrew and Erna Viterbi Faculty of Electrical and Computer Engineering (ECE) at the Technion -- Israel Institute of Technology. The authors wish to thank Omer Malka of the High-Frequency Integrated Circuits (HFIC) Lab (ECE, Technion), Maxim Meltsin of the Communication Lab (ECE, Technion), and Menashe Moritz (ECE, Technion) for their valuable technical assistance. They also thank Prof.\ Mordechai (Moti) Segev, Prof.\ Guy Bartal, and Amit Kam of the Technion for fruitful discussions.

\section*{Author contribution}
A.\ E.\ conceived the project and supervised all its aspects. A.\ S.\ derived the theory, conducted the measurements, and analyzed the results. B.-Z.\ J., I.\ V., and D.\ D.\ conceptualized, designed, and fabricated the measured device. A.\ S.\ and A.\ E. prepared and edited the manuscript. All the authors generally contributed to this work and participated in the preparation of the manuscript.

\section*{Competing interests}
All authors declare no competing interests.

\end{document}